\documentclass[11pt]{article}
\usepackage{hyperref}
\pdfoutput=1
\begin{document}
\title{Liquid ring}
\author{Seungho Kim and Ho-Young Kim\\
\\\vspace{6pt} School of Mechanical and Aerospace Engineering,
\\ Seoul National University, Seoul 151-744, South Korea}
\maketitle
\begin{abstract}
This article describes the fluid dynamics video for “Liquid ring” presented at the 64th Annual Meeting of the APS Division of Fluid Dynamics in Baltimore, Maryland, November 20-22, 2011.
\end{abstract}
\section{Introduction}
While the motion of liquids on substrates of homogeneous wettability has been studied extensively, the understanding of the behavior of liquid on substrates of non-homogeneous wettability is far from complete yet. Here we present the high-speed imaging results of the motion of water drops on a superhydrophilic annulus surrounded by a superhydrophobic area. It is found that when a drop having a volume lower than a critical value cannot maintain a spherical cap shape on the pattern but is ejected from the inner superhydrophobic circle while leaving a liquid trail on the hydrophilic annulus, which we term a ``liquid ring". These intriguing dynamics of the spontaneous morphological transition can be triggered by the collision of a water drop with the patterned surface (so that a small volume of drop can be spread to a large extent) or by letting the drop evaporate on the surface (so that its volume can reduce below the critical value). The transition is because the interfacial energy of the spherical cap is higher than the combination of the liquid ring and an isolated drop. The energy difference is consumed as a liquid drop is repelled from the inner hydrophobic pattern, which is so energetic that the drop bounces off the surface starting from a stationary spherical-cap configuration. 

%
\end{document}